\documentclass[%
 reprint,
 amsmath,amssymb,
 aps,
]{revtex4-1}

\usepackage{graphicx}
\usepackage{dcolumn}
\usepackage{bm}

\begin{document}

\preprint{APS/123-QED}
\title{The evolution of carrying capacity in constrained and expanding tumour cell populations}

\author{Philip Gerlee}
\email{gerlee@chalmers.se}
\affiliation{Mathematical Sciences, Chalmers University of Technology and G\"{o}teborg University}
\author{Alexander R.A. Anderson}%
 \affiliation{%
 Integrated Mathematical Oncology, Moffitt Cancer Center\\
 12902 Magnolia Drive Tampa, FL 33612
}%

%
%

\date{\today}

\begin{abstract}
Cancer cells are known to modify their micro-environment such that it can sustain a larger population, or, in ecological terms, they construct a niche which increases the carrying capacity of the population. It has however been argued that niche construction, which benefits all cells in the tumour, would be selected against since cheaters could reap the benefits without paying the cost. We have investigated the impact of niche specificity on tumour evolution using an individual based model of breast tumour growth, in which the carrying capacity of each cell consists of two components: an intrinsic, subclone-specific part and a contribution from all neighbouring cells. Analysis of the model shows that the ability of a mutant to invade a resident population depends strongly on the specificity. When specificity is low selection is mostly on growth rate, while high specificity shifts selection towards increased carrying capacity. Further, we show that the long-term evolution of the system can be predicted using adaptive dynamics. By comparing the results from a spatially structured vs.\ well-mixed population we show that spatial structure restores selection for carrying capacity even at zero specificity, which a poses solution to the niche construction dilemma. Lastly, we show that an expanding population exhibits spatially variable selection pressure, where cells at the leading edge exhibit higher growth rate and lower carrying capacity than those at the centre of the tumour. 

\end{abstract}

\pacs{Valid PACS appear here}
\maketitle
\section{Introduction}
A key defining feature of cancer cells is their upregulated rate of cell division, but, as we have learnt during the last couple of decades, tumour growth is dependent on a number of cellular characteristics that together drive the expansion of the lesion \cite{hanahan2000}. These traits are acquired in a process of Darwinian evolution where subclones compete for limited space and resources. One such trait is the ability of cancer cells to thrive at cellular densities, which are considerably higher than those found in normal organs \cite{neri1982}. In other words the cancer cells have a higher carrying capacity than normal cells and this property is known to correlate with aggressiveness \cite{neri1982}. 

{Progression of ductal carcinoma in situ (early stage breast cancer) provides an illuminating example. In the healthy state the duct is hollow, and lined with a single layer of epithelial cells that attach to the basement membrane. The integrity of the tissue is largely driven by homeostatic control through the balance of birth and death and regulation of growth through contact inhibition. At the onset of disease the cancer cells are confined to the existing layer where they compete with normal cells. Subsequent mutations makes it possible for the cancer cells to grow further into the duct. At the final stage of ductal carcinoma in situ the duct is filled with cancer cells, and in order for growth to continue the cancer cells breach the basement membrane and invade the surrounding tissue. This temporal progression shows how the carrying capacity increases as the disease progresses. In, fact it is for this very reason that breast screening is so effective, breast palpation (physical examination) attempts to feel for denser lumps as a key indicator of the existence of a tumour nodule. Similarly, mammogram screening highlights significant changes in breast tissue density.}

Increased carrying capacity is achieved through a number of mechanisms of which some are internal to the cell and others, that are brought about by changing the microenvironment of the cell. For example a subclone that is able to divide without being in contact with the basement membrane \cite{frisch1994}, or that can withstand high levels of acidity \cite{gatenby2004}, can grow in environments where other cells would perish. The problem of high acidity could also be tackled by  releasing endothelial growth factors that stimulate the formation of blood vessels, whose presence would reduce the acidity and increase the oxygen tension. This latter strategy would however benefit not only the cells that produce the growth factor, but all cancer cells in vicinity of the subclone. Another example of an unspecific response is the release of autocrine factors by cancer cells which increase their ability to divide at high cellular densities. This phenomenon was recently investigated by Archetti et al.\ \cite{Archetti2015} in the context of cancer cells that produce the growth factor IGF-II. They could show that producers were favoured at low serum concentrations (harsh conditions), and that intermediate concentrations gave rise to co-existence between producers and non-producers. That study was carried out \textit{in vitro}, but similar dynamics have been observed in a mouse model of tumour growth, where polyclonal tumours grew faster due to the production of growth factors by a minority subclone \cite{marusyk2014}. {The diffusible growth factor spread throughout the tumour and affected not only the producers, but a considerable fraction of the tumour and possibly the surrounding stroma.}


The totality of changes to the microenvironment that a subclone brings about, be it through autocrine growth factors, increased vessel density or the attraction of immune cells, can be viewed as a niche constructed by the subclone itself \cite{odling2003,kareva2011}, a phenomenon also known as ecosystem engineering \cite{yang2014}. This is a well-studied phenomenon in evolutionary biology and it has been established that many species increase their carrying capacity through niche construction, whereby the organisms alter their environment in such a way that it can sustain a higher number of individuals \cite{Erwin:2008p2054,Esperanza:2012p2086}. For example the ant species \textit{Myrmelachista schumanni} favours the growth of the tree \textit{Duroia hirsuta}, in which it nests, by producing formic acid that is detrimental to other plants, and this in turn leads to more nesting sites for the ants \cite{Frederickson:2007p2044}. Another example is the production of biofilm by certain bacterial species. This protective structure formed by polysaccharides is known to increase antibiotic resistance, but the chemical composition of the biofilm also influences colony size \cite{Ramos:2010p2058}, and hence the carrying capacity of the bacterial strain. 

Just as in the case of growth factor production by cancer cells, these situations allow for the possibility of cheating or free-riding on the strain or subclone that facilitates the increased carrying capacity. The ability to do so largely depends on the specificity of the niche construction activity. If the modification of the niche is highly specific to the genotype that generates it, then most likely it is harder for other genotypes to exploit it, whereas a more general modification is easier to free-ride on.  A natural question that arises in this context is how the specificity of the niche construction activity alters the evolutionary dynamics of the system. Another aspect, highly relevant in the case of cancer, is the impact niche construction has on a spatially expanding tumour population. 

We have investigated the evolution of carrying capacity in the context of breast cancer growth, since these tumours originate in a confined anatomical structure, a duct, in which the cancer cells outcompete the normal epithelial cells, increase the cell density and eventually break the basement membrane to become an invasive tumour. The model we have studied is however quite general and the conclusions drawn from it are applicable to any biological system where organisms increase their carrying capacity by means of a more or less specific niche construction. 


\section{Individual-based model}
The mammary ducts form a tubelike structure, which in the normal state are lined with a single layer of epithelial cells. In the early stages of breast cancer, tumour cells are still confined to the duct, but exhibit a higher cellular density \cite{tavassoli1998}. Eventually the basement membrane that encapsulates the ducts is breached and the tumour becomes invasive, expanding into the surrounding tissue. 

To model this we consider a one-dimensional spatially discrete individual-based model where each lattice site can either be occupied by a single normal cell or one or more cancer cells. The aim of our model is to capture the evolutionary dynamics of a tumour that contains a collection of subclones that differ in their growth rate $r$ and the ability to survive at high cellular densities (i.e.\ carrying capacity $k$).
The ability to survive at high densities is assumed to depend on their ability to change the local microenvironment and construct a niche, in a more or less specific manner.

In the model the specificity of the niche construction is controlled via a parameter $\gamma \in [0,1]$, where $\gamma = 0$ represents the case where all beneficial changes to the environment are shared equally between cells at the same spatial location (no specificity, i.e.\ all cancer cells at a site share the same carrying capacity given by the mean carrying capacity of the cells at that site), and $\gamma = 1$ corresponds to no sharing of the niche (the maximal density a cell can endure is determined by the cell's intrinsic carrying capacity). For intermediate values of $\gamma$ the effective carrying capacity experienced by a cell depends both on its intrinsic ability and the context it is in. 

\subsection{Model implementation}
The individual-based model of tumour growth is discrete in both time and space. The computational domain consist of $L$ lattice points, and each point is either empty (or equivalently contains a normal cell) or inhabited by one or more cancer cells. The cancer cells differ in their ability to divide and withstand high cellular densities, and this is captured by two parameters: the growth rate $r$ and the  intrinsic carrying capacity $k$. Each cell experiences an effective carrying capacity $K_{i,l}$, which depends both on the intrinsic carrying capacity of cell $i$ and on the spatial location $l$, according to
\begin{equation}
K_{i,l} = \gamma k_i + (1-\gamma)\bar{K}_l
\end{equation}
where $\bar{K}_l$ is the average carrying capacity of the cells at location $l$. At each time step $\Delta t$ all the cancer cells are updated in random order according to the following scheme:

With probability $r\Delta t $ the cell divides and with probability $1-N_l/K_{i,l}$, where $N_l$ is the number of cells at lattice site $l$, the daughter cell is placed at the same location as the parent. If this fails the daughter cell is either placed at site $l+1$ with probability $1-N_{l+1}/K_{i,l+1}$ or at $l-1$ with corresponding probability. If all these alternatives fail the daughter cell dies immediately. After cell division has been attempted the parent cell dies with probability $\delta \Delta t$.

We use non-flux boundary conditions such that the cells at $i=1$ and $i=L$ only have one neighbouring lattice point. As an initial condition we fill the domain with cells that all have $r=k=1$.

If cell division is successful each trait of the daughter cell is mutated separately with probability $\mu = 10^{-2}$, and the mutant's parameters are set to: $k' = k + \varepsilon_k$ and $r'=r+\varepsilon_r$ respectively, where $(k,r)$ are the parameters of the parent cell and $\varepsilon_{k,r}$ are drawn from normal distributions with mean zero and variance $\sigma_{k,r} = 1 $. {The fact that mutations to the traits are considered separately, together with the small mutation rate, implies that almost all mutants ($\approx 99.99\%$) will differ in one trait only (compared to the parent cell). The two traits can thus evolve independently and the model, initially, does not contain any trade-off between $r$ and $k$.} {It is natural to assume that $k$ and $r$ remain positive, and we therefore consider any mutant with negative $k$ or $r$ as unviable.}

{The rate of mutation of cancer cells is considerably smaller than our choice of $\mu=10^{-2}$ (on the order of $10^{-8}$ per base pair) \cite{Tomlinson1996}, but here we are  dealing with traits that are potentially controlled by hundreds of genes, whose impact on the phenotype is determined by the highly complex genotype-phenotype map. Our choice of $\mu$ and $\sigma_{k,r}$ are therefore rough estimates, chosen partly with computational expediency in mind. The results presented remain qualitatively similar for smaller values of $\mu$ and $\sigma_{k,r}$ (data not shown).}


\section{Results}
To capture the two phases of intra-ductal and invasive growth we have analysed the impact of the specificity $\gamma$ on the evolutionary dynamics of the system in both a domain of fixed size and in a spatially expanding population. The former case is amenable to analysis and we formulate a mean-field model, which predicts both the short-term and long-term evolutionary dynamics of the fixed size system. 

\subsection{Fixed domain size.} 
\begin{figure}[tbh]
\begin{center}
\includegraphics[width=9cm]{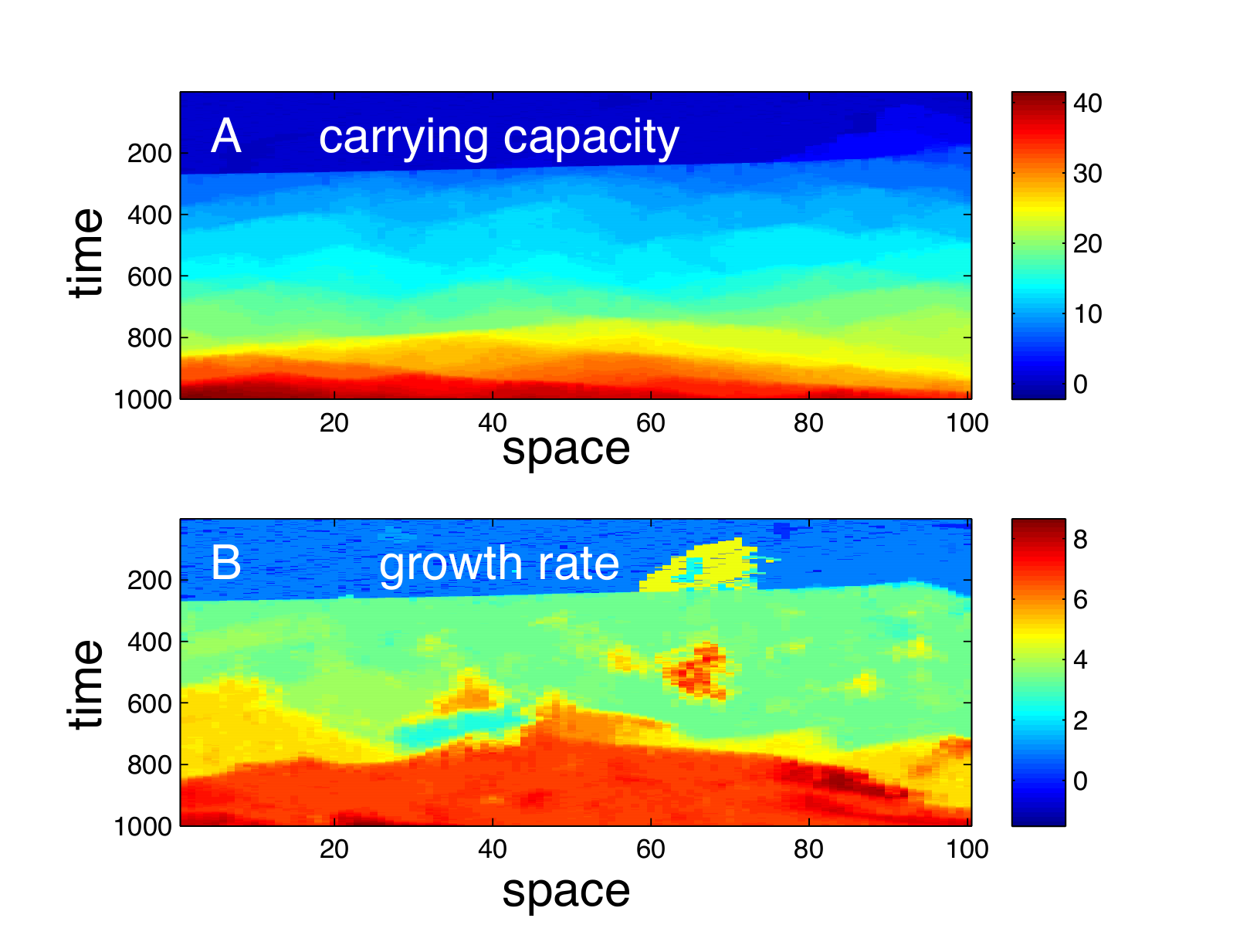}
\caption{\label{fig:fig1}The time evolution of the (a) carrying capacity and (b) growth rate as a function of time and space. The simulation is started with a homogeneous population with $(k,r)=(1,1)$. The specificity is set to $\gamma=1/2$, the death rate to $\delta = 0.1$ and the mutation rate is $\mu = 10^{-2}$.} 
\end{center}
\end{figure}

We initiate the model with a homogeneous population of cells with intrinsic carrying capacity $k=1$ and growth rate $r=1$. The system is simulated with death rate $\delta = 0.1$ on a domain of $L=100$ lattice sites. The result of a simulation where the niche specificity is set to an intermediary value ($\gamma = 1/2$) is shown in figure \ref{fig:fig1}. The two panels show 
the carrying capacity (A) and growth rate (B) as a function of space and time. We observe successive selective sweeps of clones that have higher carrying capacity. The growth rate also increases over time, but here selection is less persistent and some of the mutant clones with higher $r$ are driven to extinction (e.g. at $t\approx 500$ and $l=65$). Note that although the domain is of fixed size the population size increases in step with the increase in carrying capacity. 

\begin{figure}[tbh]
\begin{center}
\includegraphics[width=9cm]{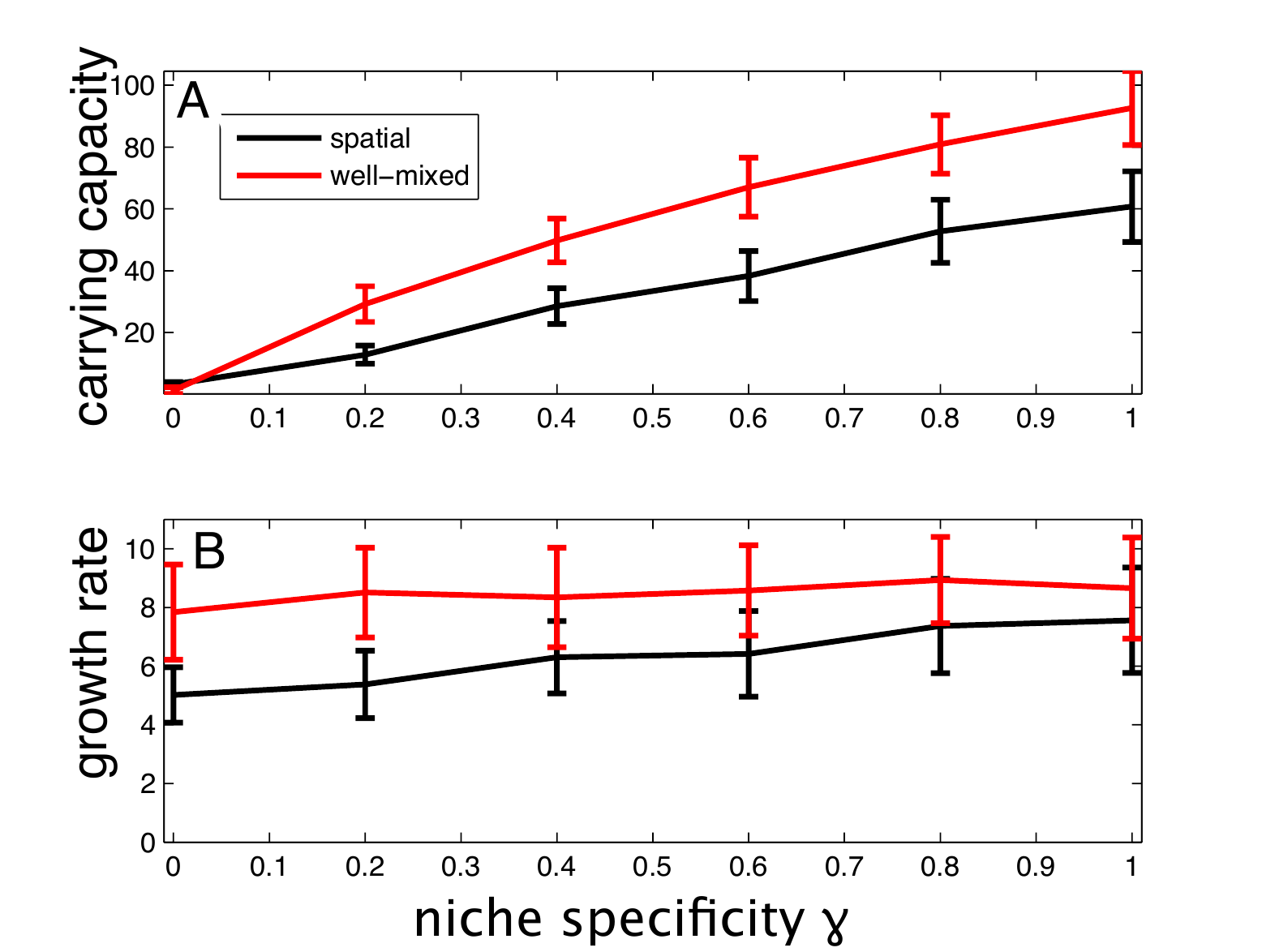}
\caption{\label{fig:fig2}The average phenotype in the population after 1000 time steps (generations of the wild-type population) as a function of the niche specificity $\gamma$. (a) shows the average carrying capacity $\bar{k}$ and (b) shows the average growth rate $\bar{r}$. The results are averaged across 100 simulations and the error bars corresponds to one standard deviation. {The difference between the carrying capacity and growth rate in the spatial and well-mixed case are statistically significant for all values of $\gamma$ (rank-sum test $p < 0.0023$).} Each simulation is started a homogeneous population with $(k,r)=(1,1)$. The death rate is set to $\delta = 0.1$ and the mutation rate is $\mu = 10^{-2}$.} 
\end{center}
\end{figure}

The selective pressure on the carrying capacity and the growth rate is explored further in figure \ref{fig:fig2}, which shows the mean carrying capacity and growth rate in the entire population as a function of the niche specificity $\gamma$. In order to investigate the impact of space we have also performed identical simulations with a model where the location of the daughter cell is chosen at random (i.e.\ a well-mixed population). The results are averaged across 100 simulation for each value of $\gamma$.

From this figure it is clear that the niche specificity has a strong impact on the evolutionary dynamics of the system. At low specificities, the average carrying capacity changes little from the initial value of $k=1$, while for $\gamma=1$ selection  leads to an almost sixty-fold increase in carrying capacity. The growth rate on the other hand exhibits almost constant selection and only a slight increase is evident. The results from the well-mixed version of the model are similar, although in this case selection for $k$ is even stronger (except for $\gamma=0$), and 
 $r$ is essentially independent of $\gamma$. {The difference between the average carrying capacity and growth rate obtained in the spatial and well-mixed case are statistically significant for all values of $\gamma$ (rank-sum test $p < 0.0023$).} 

\subsection{Mean-field model.}
In order to better understand the results obtained from the individual-based model and to investigate the impact of model parameters we have derived a mean-field model (see SI for details). The model describes the average number of cells $x_i$ of type $i$ per site, which changes according to: 
\begin{equation}\label{eq:sys}
\frac{dx_i}{dt} = r_ix_i \left( 1-\frac{x_T}{{K}_i(\bm{x})} \right) - \delta x_i
\end{equation} 
where $r_i >0$ is the growth rate of type $i$, $x_T = \sum_i x_i$ is the average number of cells per site, $K_i(\bm{x})$ is a subclone specific carrying capacity that depends on the current subclone composition $\bm{x} = (x_1,x_2,...,x_n)$ and $\delta > 0$ is a density independent death rate assumed equal for all subclones. The carrying capacity of type $i$ is given by

\begin{equation}
K_i(\bm{x})  = \gamma k_i + \frac{1- \gamma}{x_T}\sum_{j=1}^n k_j x_j
\end{equation}
where $\gamma$ again controls the specificity of the carrying capacity. Please note that in the presence of a single subclone the system reduces to the standard logistic equation for all values of $\gamma$. 

\subsection{Stability analysis.}
The evolutionary dynamics of the system in the short-term depends on the ability of a mutant to spread and invade a resident population. This question can be resolved by analysing the stability of a resident subclone when a mutant with a different phenotype is introduced into the population. 

To this end we consider the situation where only $n=2$ subclones are present (a resident and a mutant), in which case the system of equations simplifies to
\begin{eqnarray}
\frac{dx_1}{dt} &= r_1x_1 \left( 1-\frac{x_1+x_2}{K_1(x_1,x_2)} \right) - \delta x_1  \nonumber \\
\frac{dx_2}{dt} &= r_2x_2 \left( 1-\frac{x_1+x_2}{K_2(x_1,x_2)} \right) - \delta x_2,
\end{eqnarray} 
where 
\begin{equation}
K_{1,2}(x_1,x_2)  = \gamma k_{1,2} + \frac{1-\gamma}{x_1 + x_2}(k_1 x_1 + k_2 x_2).
\end{equation}
{We now assume (without loss of generality) that subclone $1$ is the resident and $2$ is the mutant, and also that $k_1 > 0 $ and $r_1 > \delta$, i.e.\ the resident has a positive carrying capacity and its growth rate is larger than its death rate.}

The steady-state, corresponding to a monomorphic population, is given by 
\begin{equation} \label{eq:eq}
\bm{x}^\star = (x_1,x_2) = (k_1(1-\delta/r_1),0)
\end{equation}
where $k_1(1-\delta/r_1) > 0$ due to the above assumptions.

The mutant can invade the resident population if the steady-state is unstable. This is determined by the eigenvalues of the linearised system, which are given by $\lambda_1 = \delta - r_1$ and 
\begin{equation}\label{eq:S}
S(\gamma) = r_2 \left(1- \frac{k_1}{\gamma k_2 + (1-\gamma) k_1}(1-\delta/r_1) \right) - \delta.
\end{equation}
Since $\delta - r_1 < 0$ (by the above assumption), the ability of a mutant to invade depends on the sign of $S(\gamma)$, i.e. the mutant can invade whenever $S > 0$.
We start by investigating the extremes, $\gamma = 0$ and 1. In the case of minimal specificity ($\gamma=0$) we have 
\begin{equation}
S(0) = \delta \left(\frac{r_2}{r_1} - 1\right),
\end{equation}
which implies that the mutant can invade if $\delta > 0 $ and $r_2 > r_1$. 
For the case with maximal specificity ($\gamma=1$) we have
\begin{equation}
S(1) = r_2 \left(1- \frac{k_1}{k_2}(1-\delta/r_1) \right) - \delta.
\end{equation}
If the mutant differs in carrying capacity and hence $k_1 \neq k_2$, and in addition $\delta \ll r_1$, we have $S(1) \approx r_2 (1-k_1/k_2)$, which implies that $S > 0$ if and only if $k_2 > k_1$, i.e.\ the mutant can invade if it has a larger carrying capacity than the resident. On the other hand if the mutant differs in growth rate, and hence $k_1 = k_2$, then $S(1) = S(0)$ (in fact $S$ becomes independent of $\gamma$) and again only mutants with a larger growth rate can invade.

\begin{figure}[!htb]
\begin{center}
\includegraphics[width=8cm]{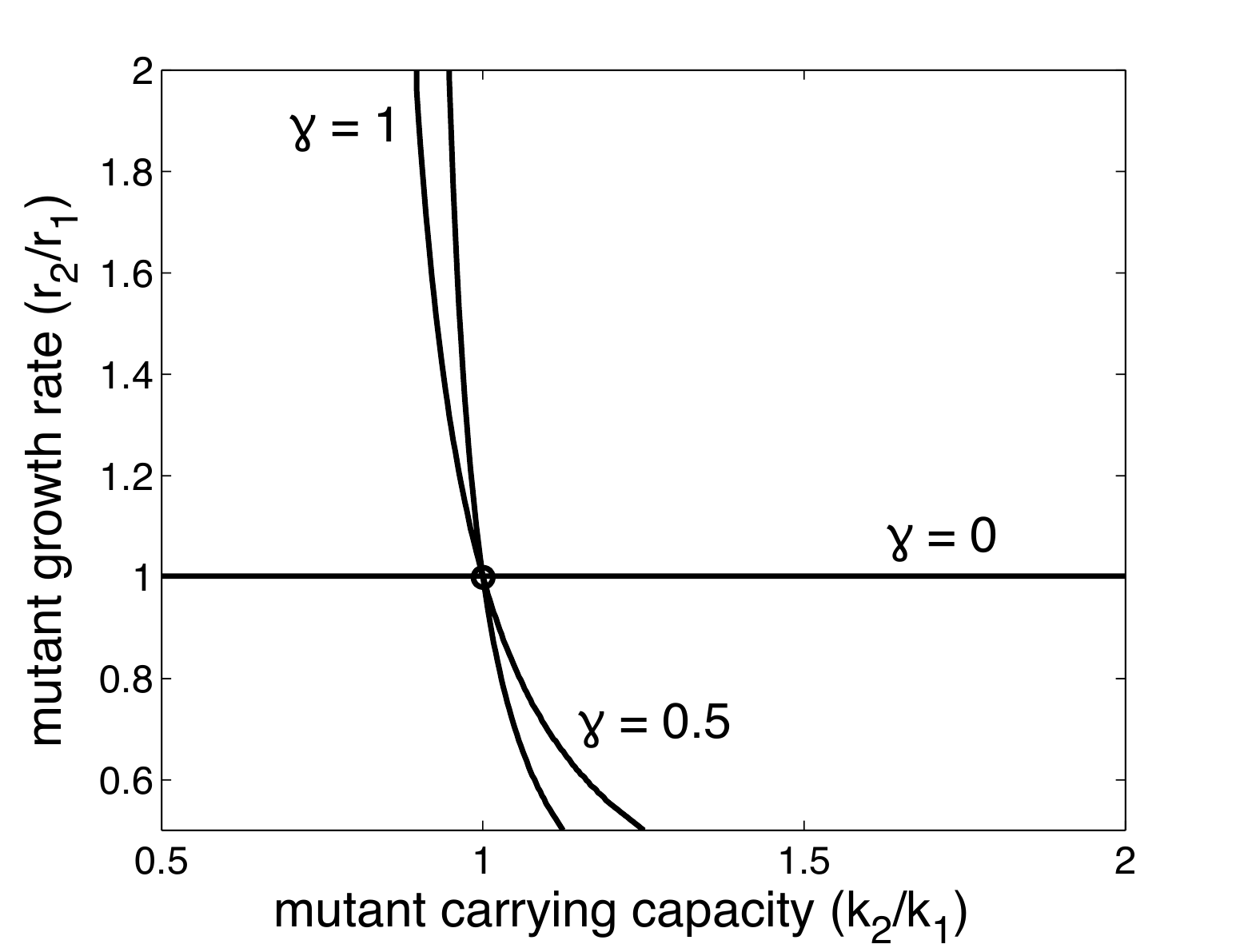}
\caption{\label{fig:fig4}Invasion as a function of mutant $r$ and $k$. The region above (and to the right) of the curves correspond to the subset of mutant characteristics, in terms of the relative growth rate ($r_2/r_1$) and relative carrying capacity ($k_2/k_1$), for which a mutant can invade. In the local case ($\gamma = 1$) dependence is almost exclusively in the $k$-direction, whereas in the global case ($\gamma=0$) only the mutant growth rate affects the ability to invade. As is evident from the curve corresponding to $\gamma = 0.5$, the transition to a nearly vertical curve occurs for small values of $\gamma$. The death rate was set to $\delta = r_1/10$.} 
\end{center}
\end{figure}

In order to get a better understanding of the impact of the specificity we plot the curve $S(k_2,r_2)=0$ in the $(r,k)$-parameter space for three different choices of $\gamma$ (see figure \ref{fig:fig4}).
The regions above (and to the right) of the curves correspond to the subset of mutant characteristics for which a mutant can invade. In the global case ($\gamma=0$) the ability to invade depends only on the mutant growth rate whereas in the local case ($\gamma = 1$) dependence is almost exclusively in the $k$-direction. {As is evident from the intermediate case ($\gamma = 0.5$), the curve $S=0$ shifts from horizontal to nearly vertical already for small values of $\gamma$. This shift occurs for even smaller $\gamma$ if the death rate $\delta$ is reduced.}

{This analytical result helps us understand the results from the IB-model. From the discussion above we know that if the mutant differs in growth rate then it is able to invade independent of the value of $\gamma$. This explains why selection for growth rate is approximately constant for all $\gamma$ (see fig.\  \ref{fig:fig2}). On the other hand, the growth rate \eqref{eq:S} of a mutant with $k_2 > k_1$ is an increasing function of $\gamma$, which implies that such mutants are more likely to reach fixation at higher $\gamma$. This explains why selection for carrying capacity increases with $\gamma$ (see fig.\  \ref{fig:fig2}).}


\subsection{Adaptive dynamics.}
The above analysis explains the local evolutionary dynamics, but is unable to predict the long-term dynamics of the system. In order to investigate this question we make use of the adaptive dynamics framework \cite{brannstrom2013}. Assuming that there is a separation between the time scale of population dynamics and evolutionary dynamics and that mutations give rise to small changes in phenotype this framework can predict the course of evolution, and in particular identify equilibrium points of the evolutionary dynamics. {Initially we analyse the case were $r$ and $k$ evolve without any constraints, and then analyse the situation where a trade-off between growth rate and carrying capacity is assumed.}

\begin{figure}[!htb]
\begin{center}
\includegraphics[width=8cm]{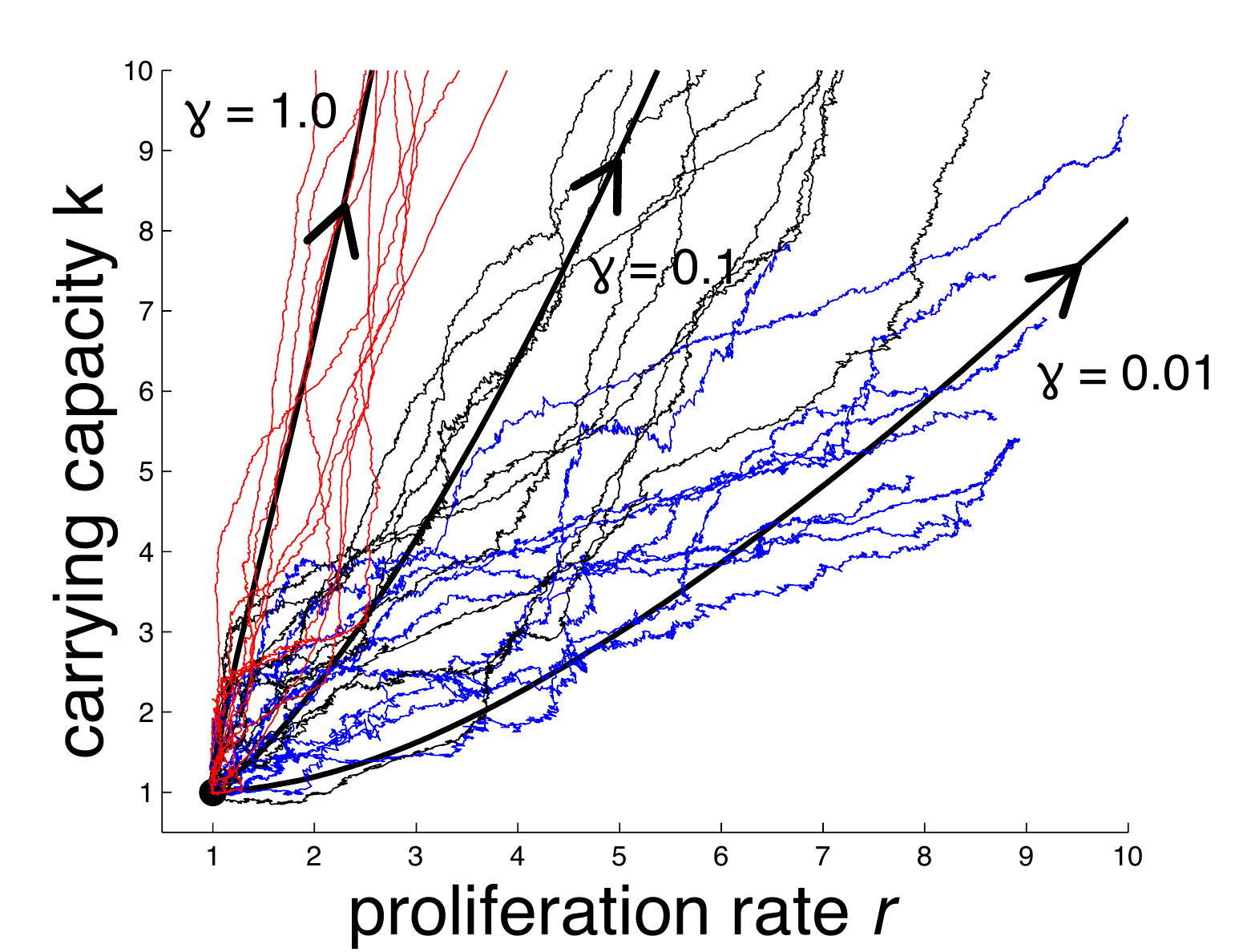}
\caption{\label{fig:selection} Solution of the canonical equation \eqref{eq:canon} of trait evolution for $\gamma = 0.01,\ 0.1$ and $0.1$ (solid black curves), and trait substitution sequences from the individual-based model (jagged curves).}
\end{center}
\end{figure}

If we assume that there is no covariance between the traits then the canonical equation for our system, which describes the mean evolutionary trajectory, can be written

\begin{eqnarray} \label{eq:canon} 
\frac{dr}{dt} &= \frac{\mu \sigma_r^2 \delta}{2r} k (1-\delta/r) \nonumber \\
\frac{dk}{dt} &=   \frac{\mu \sigma_k^2 \gamma}{2r} (r-\delta)^2,
\end{eqnarray} 
where $\sigma_{k,r}^2$ is the variance of the mutational steps (see SI for details). Figure \ref{fig:selection} shows the solution of the canonical equation with initial condition $(r,k)(t=0) = (1,1)$ for $\gamma = 0.01,\ 0.1$ and $1.0$ (solid black curves). Shown here is also ten trait substitution sequences from the IB-model for each value of $\gamma$, that approximately follow the solutions of the canonical equation. 

As expected, small values of $\gamma$ lead to selection in both the $k$- and $r$-direction, and increasing $\gamma$ shifts selection further in the $k$-direction. {The small domain size ($L=100$) leads to a large amount of drift (in particular for small $\gamma$ where population size remains small), but for larger system sizes there is, as expected, less drift (fig.\ S2). Other factors that cause the simulation results to deviate from the canonical equation are the spatial structure in the population, the relatively large mutation rate, which leads to an overlap between ecological and evolutionary dynamics, and the non-infinitesimal mutational steps. In total this implies that not even the average trait substitution curve of the IB-model follows the solution of the canonical equation exactly (fig.\ S3). }

{The canonical equation can still be used to analyse possible evolutionary endpoints.} A rest point of the evolutionary dynamics, known as a singular strategy, is a point $(k^\star,r^\star)$ such that the right hand side of eq.\ \eqref{eq:canon} vanishes, i.e. the selection gradient is 0. We can see from \eqref{eq:canon} that the line defined by $\{(r,k); r = \delta \ {\rm and\ } k \geq 0 \}$ contains all singular strategies of the canonical equation. Linearisation of the system shows that this is an unstable manifold, and therefore not reachable by the evolutionary dynamics \cite{leimar2005}, and in addition is excluded as an initial condition since we have assumed that $r > \delta$. We can therefore conclude that if there is no trade-off between growth rate and carrying capacity the evolutionary dynamics have no attainable singular strategies, and we observe run-away selection in both $k$ and $r$ along a trajectory determined by the canonical equation \eqref{eq:canon}.

We will now proceed to investigate the effect of a strict trade-off between $r$ and $k$. This is realised by letting $r=f(k)$, here we specifically analyse the case $f(k)=r_0 e^{-ak}$, where $r_0 > \delta$. In this situation the selection gradient is given by (see SI for details)
\begin{equation}\label{eq:1dgrad}
D(k)= \frac{\gamma}{k} (r_0 e^{-ak} - \delta) - \delta a.
\end{equation}
For the global case of $\gamma = 0$ we have $D(k) = -\delta a < 0$. Selection is always for smaller $k$ (and hence larger $r$). No non-zero singular strategy exists, and we expect the carrying capacity to evolve towards $k=0$.

For $\gamma > 0$ we have $D(k) = \gamma (r_0e^{-ak} - \delta)/k - \delta a$, which is zero for
\begin{equation}\label{eq:sol}
k^{\star} = \frac{W(\frac{e^\gamma \gamma r_0}{\delta}) - \gamma}{a}
\end{equation}
where $W(\cdot)$ is the Lambert $W$ function. 
This singular strategy is both evolutionary and convergence stable for all $\gamma \in [0,1]$, and thus an evolutionary end-point of the system (see SI for details).


This result is only exact in the limit of a large population size, infinitesimal mutational steps, and a separation of ecological and evolutionary time scales. In order to test its validity under less stringent and more realistic conditions we compare the prediction from the adaptive dynamics with the long-term dynamics of the IB-model. The trade-off is implemented by only considering mutations to the carrying capacity and letting $r_i=r_0 e^{-ak_i}$ for all subclones. 

\begin{figure}[!htb]
\begin{center}
\includegraphics[width=8cm]{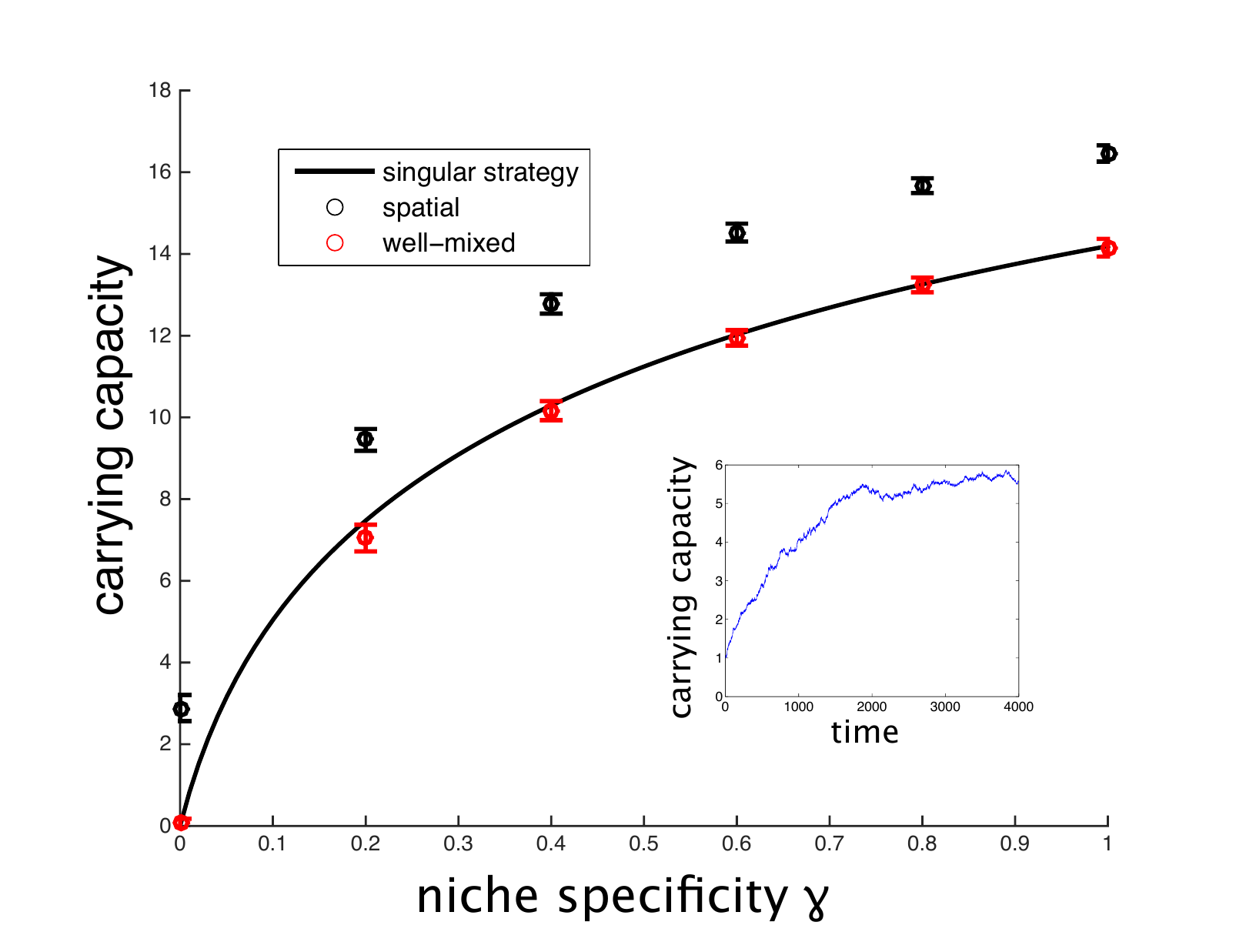}
\caption{\label{fig:fig6} The solid line shows the singular strategy $k^{\star}$ obtained from \eqref{eq:sol}, while the circles show the average carrying capacity in the IB-model after $T=5000$ time steps in the spatial (black) and well-mixed (red) cases. Error bars correspond to one standard deviation. The inset shows the time evolution of the mean carrying capacity in a single run of the individual-based model for $\gamma=0.1$. The trade-off between growth rate and carrying capacity is set to $r=r_0 e^{-ak}$, where $r_0 = 1$ and $a=0.1$.} 
\end{center}
\end{figure}

Each simulation was initiated with a homogeneous population with $(k,r) = (1,r_0 e^{-a})$, where $a=0.1$ and $r_0=1$, on a domain of $L=100$. The average carrying capacity in the population after 5000 time steps, averaged across 50 simulations, can be seen in fig.\ \ref{fig:fig6}. The results obtained using adaptive dynamics agrees well with those from the well-mixed model, but underestimates the carrying capacity in the spatial case. 

A potential source of error is the spatial organisation of subclones, which leads to higher than expected (compared to the well-mixed scenario) interactions between cells with identical carrying capacity. In fact the probability (when the system is stationary at the singular strategy) that two cells that reside on the same grid point belong to the same subclone is $p = 0.9$ for $\gamma = 0$ (and $p=0.5$ for $\gamma = 1$) and decreases exponentially as a function of the distance between the cells (see fig.\ S1).


\subsection{Expanding population}
We have so far been concerned with analysing the dynamics of a population evolving on a domain of fixed size. The population size has not been limited, but any increase in size in this scenario is due to changes in the carrying capacity of the cells. This is reminiscent of the early stages of breast tumour growth when the lesion is confined within the duct basement membrane. At high enough cellular densities the membrane is however breached and the tumour becomes invasive and enters into a new phase of expanding growth. In order to examine this latter scenario we simulate the model on a larger lattice ($L=400$). We initiate the simulation with a single cell with $(r,k)=(1,1)$ at the left side of the domain ($l=1$) and terminate the simulation when the rightmost site at $l=L$ has become occupied (approximately after 500 time steps). 

\begin{figure}[!tbh]
\begin{center}
\includegraphics[width=8cm]{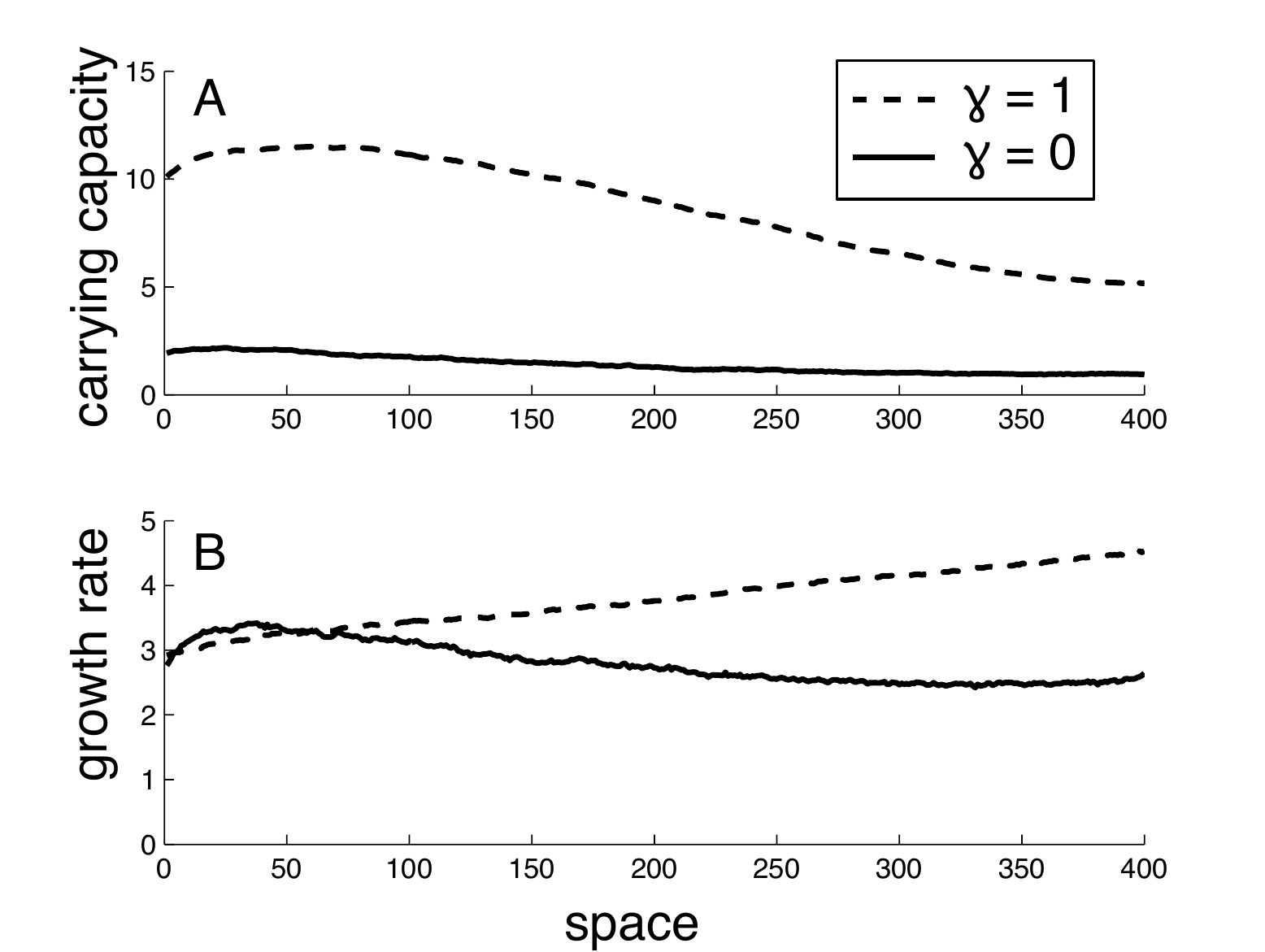}
\caption{\label{fig:spat}Spatial phenotypic heterogeneity in an expanding population for $\gamma = 0$ and $1$. The carrying capacity is maximal close to the centre of the population, while the growth rate is highest at the boundary. The curves were obtained by averaging the results across 1000 simulations.} 
\end{center}
\end{figure}

Figure \ref{fig:spat} shows the mean growth rate and carrying capacity as a function of space and the results are averaged across 1000 different simulations. For $\gamma = 1$ we observe spatial heterogeneity, where the carrying capacity is largest close to the origin of the population, while the growth rate exhibits an opposite trend, and reaches its maximum value at the boundary of the population. Also, it is worth noting that for $\gamma = 1$ the ratio of $k/r$ at the boundary ($L=400$) is $\approx 1$, while a fixed population that has evolved for the same amount of time we find a ratio of approximately $k/r \approx 4.7$. For low specificity the spatial structure is less pronounced, and for both $k$ and $r$ there is a slight increase close to the origin. 

These results suggests that the evolutionary dynamics of an expanding population are quite different from that of a fixed one, and hence that the selection pressures experienced by early tumours might be reversed in later stages of growth.  

\section{Discussion}

We  presented a simple model of tumour growth, where niche-construction (e.g.\ growth factor secretion) is assumed to affect the carrying capacity of not only the cell itself, but also the carrying capacity of adjacent cells. This effect is controlled by the specificity parameter $\gamma$, which we have shown has a strong impact on the dynamics of the model, and our results suggest that only a small  degree of specificity in niche construction is sufficient for a persistent selection for increased carrying capacities. 

In the case of zero specificity the stability analysis carried out on the mean-field model (see fig.\ \ref{fig:fig4}) shows that there is only selection for higher growth rate. This is also reflected in the canonical equation \eqref{eq:canon}, where $\gamma = 0$ implies no evolutionary change in the $k$-direction. However, even for small values of $\gamma$ selection is shifted towards carrying capacity (e.g.\ the curve for $\gamma = 0.5$ in fig.\ \ref{fig:fig4}). The increase in carrying capacity in turn leads to a larger population, which accelerates the evolutionary process even further (note the dependence on $k$ in $\frac{dr}{dt}$ of the canonical equation \eqref{eq:canon}). 

When no trade-off is present between growth rate and carrying capacity, the selection for higher $k$ and $r$ continues in a runaway fashion and there are no stable strategies. This can be changed by introducing a trade-off function, $r=f(k)$, and we have shown that in this case the adaptive dynamics framework can predict the long-term evolutionary dynamics of the system (see fig.\ \ref{fig:fig6}), in particular when the system is well-mixed. 

Lastly we have shown that a population that is expanding spatially is subject to different dynamics and that the selection pressure depends on space, where cells close to the leading edge divide faster and have a lower carrying capacity than the cells at the origin.

The relationship between selection for carrying capacity and growth rate, usually termed $r/K$-selection has a long history in evolutionary biology dating back to the seminal work of MacArthur in the 1960s \cite{MacArthur:1962p2059}. Although the paradigm has largely been replaced by more detailed studies of life history evolution, it has provided important insight into the evolutionary dynamics of density limited populations. For example it has been shown that populations that evolve in stable and mild environments will experience selection for high $K$-values, while populations in harsh seasonal environments will evolve towards higher $r$ \cite{Roughgarden:1971p2055}. {This relationship has also been hypothesised to hold in tumours \cite{elser2007}. Tumours emerging in epithelial tissue with high cell turnover are expected to show selection for growth rate, while tumours in more stable tissues, such as the liver, are predicted to exhibit selection for carrying capacity. The results presented here suggest a novel mechanism, that can shift the balance between $r-$ and $K-$selection, which does not rely on cell turnover, but rather on the specificity of a constructed niche.} 


This mechanism has previously been explored in the context of ecosystem dynamics by Krakauer et al.\ \cite{krakauer2009}. They considered organisms of two competing species that invest in a common niche that increases the carrying capacity of both species. This leads to a construction dilemma where an invading species that invests less in the niche can invade, but the dilemma can be resolved by introducing `niche monopolisation', which is equivalent to the specificity $\gamma$ used here. In that model there is a strict linear trade-off between growth rate and niche construction, motivated by life-history choices made by the individuals. This is less appropriate for cancer cells, since trade-off between niche construction and cell division are more likely to be biochemical and energetic rather than temporal and hence not necessarily linear. In addition the model by Krakauer et al. does not have a spatial component, and therefore cannot capture the difference between a well-mixed and spatially structured population, which we have shown to exhibit different stable strategies (see fig.\ \ref{fig:fig6}).

The concept of a niche constructed for a specific subclone is related to `kin discrimination', which refers to the ability of an organism to distinguish organisms with which they are related \cite{strassmann2011}. This ability can be beneficial, especially for costly behaviours such as biofilm production, but often requires sophisticated mechanisms such as clone specific signalling. A cruder solution to the problem of identifying kin is simply to assume that your neighbours most likely are your relatives, essentially relying on proximity as a proxy for relatedness. We have shown that relatedness between proximal cells is high in the spatial version of the IB-model (see fig.\ S1). 
The spatial aggregation of clonal cells means that a cell is more likely to contribute to the carrying capacity of a cell from the same subclone, which in turn leads to an apparent increase in the specificity of the action.
This is what occurs in the spatial version of the IB-model, in contrast to the well-mixed system where daughter cells are placed randomly. The difference in dynamics between the two models (e.g.\ the higher $k^\star$ for the spatial model in fig.\ \ref{fig:fig6}) can therefore be attributed to this most basic form of kin discrimination. 

The evolution of carrying capacity has largely been disregarded in models of tumor growth, which have instead focused on increased growth rates and adaptation to adverse environmental conditions (low nutrients, cytotoxic therapy etc.). A notable exception is the work of 
Nagy \cite{nagy2005,nagy2007}, in which the niche construction dilemma in the context of cancer has been investigated. 
Using a system of ordinary differential equations that captures the dynamics of cancer cells, immature blood vessels and fully formed vessels he showed that a resident cancer cell population can be invaded by a mutant that invests less in attracting blood vessels and more in proliferation, eventually leading to `evolutionary suicide'. In a more recent study by Nagy \& Armbruster \cite{nagy2012} the model was extended by taking into account the energy management and constraints cancer cells are subject to. That model exhibited a repellant evolutionary rest point that could lead either to runaway selection for hypo- or hyperplasia. These models assume zero specificity, and in the light of our results the selection against angiogenic signalling is not surprising. The spatial organisation of subclones within the tumour might, however, be sufficient to restore selection for carrying capacity, as is suggested by the higher stable strategy in the spatial model (fig. \ref{fig:fig6}).

The dependence on niche construction specificity has important implications for our understanding of tumor growth. Naively one would expect that since many factors (e.g.\ angiogenic and autocrine signalling) that positively impact tumor growth are diffusible and hence unspecific in their impact on different subclones, they would be subject to exploitation. This would then lead to selection for higher growth rate among the cancer cells, and not towards the high carrying capacities normally observed in tumours. Our analysis has shown that very low levels of specificity are required to drive selection for carrying capacity and that even spatial organisation (proximity as a proxy for relatedness) can favour selection for higher carrying capacities. 

An important step in understanding the connection between niche construction and selection in tumours is to quantify the specificity of different mechanisms. For example, in a poorly oxygenated tumour, what is the correlation between the production of angiogenic signals and oxygenation on the scale of single cells? Do cancer cells that produce autocrine growth factors respond in the same way as non-producers do? Or, are most adaptations internal to the cell, e.g. increased tolerance to acidity and low oxygen concentration? The model also highlights the need to determine the spatial organisation of sublcones, a topic that is just starting to be explored \cite{gerlinger2012, sottoriva2013}. 

We realise that the model we have presented is a simplistic `toy-model' that omits a number of important aspects, such as a potential delay in niche construction. It is also reasonable to believe that the specificity itself is subject to evolutionary dynamics, such that each subclone could make use of a unique mechanism for altering the niche. Despite these simplifications the model provides novel insight into the impact of niche construction specificity on evolutionary dynamics, and hints at the importance of including carrying capacities as a variable when considering the evolutionary dynamics of tumour growth.

\bibliography{log.bib}
\clearpage

\section*{Supplementary material}
\subsection*{Derivation of mean-field model}
We consider a well-mixed system and our variable of interest will be the average number of cells $x_i$ of type $i$ per site. Each such cell divides at rate $r_i$ and since the offspring is placed at a random site the rate of successful division is $r_i (1-x_T/K_i)$, where $x_T = \sum_j x_j$ is the average total number of cells per site and $K_i = \gamma k_i + (1-\gamma)\bar{K}$. Here $k_i$ is the intrinsic carrying capacity of subclone $i$ and $\bar{K}$ is the average carrying capacity which is given by 
\begin{equation}
\bar{K}  = \frac{1}{x_T}\sum_{j=1}^{n} k_j x_j.
\end{equation}
Accounting for the constant death rate $\delta$, the change in $x_i$ after one time step $\Delta t$ can be written:
\begin{equation}
x_i(t + \Delta t) = x_i (t) + \Delta t \left( x_ir_i(1-\frac{x_T}{K_i(\bm{x})}) - \delta x_i \right)
\end{equation}
which in the limit $\Delta t \rightarrow 0$ turns into a coupled system of ODEs:
\begin{equation}\label{eq:sys}
\frac{dx_i}{dt} = r_ix_i \left( 1-\frac{x_T}{{K}_i(\bm{x})} \right) - \delta x_i
\end{equation} 
where $\bm{x} = (x_1,x_2,...,x_n)$ is the average number of different subclones in the population.

\subsection*{Adaptive dynamics}
The basis for the adaptive dynamics framework is $F(\bm{\hat{x}},\bm{x})$ -- the invasion fitness of a mutant with trait $\bm{\hat{x}}$ in a monomorphic population with trait $\bm{x}$. In our case the trait $\bm{x} = (r,k)$ is two-dimensional and $F$ is given by $S(\gamma)$ in equation (7) in the main text, but where $S$ now should be viewed as a function of $r$ and $k$.  From this we can deduce that $F(\bm{x},\bm{x}) = 0$, which is to be expected since any resident is in equilibrium with itself. 

From the invasion fitness one can calculate the selection gradient
\begin{equation}
\bm{D}(\bm{x})=\frac{\partial F(\bm{\hat{x}},\bm{x})}{\partial {\bm{\hat{x}}}}|_{\bm{\hat{x}}=\bm{x}}
\end{equation}
which contains information about the direction and magnitude of selection, and enters into the so called canonical equation
\begin{equation}
\frac{d\bm{x}}{dt} = m(\bm{x}) \bm{C} \cdot \bm{D}(\bm{x})
\end{equation}
where $m(\bm{x})$ is function that captures the variation in mutation frequency and $\bm{C}$ is the variance-covariance matrix of the traits considered. The canonical equation describes the mean path taken by the system in the trait space and is the deterministic limit of a stochastic trait substitution sequence \cite{dieckmann1996} -- a biased random walk that is influenced both by drift and selection. 

If we assume that there is no covariance between the traits (so that $\bm{C}$ is the identity matrix) and make use of the fact that $m(\bm{x}) = \mu N(\bm{x})/2$, the mutation rate times the equilibrium population size, where $N(r,k)=k(1-\delta/r)$, the canonical equation for our system can be written\footnote{The equilibrium population size of a monomorphic population with traits $r$ and $k$ is given by equation (6) in the main text.}

\begin{eqnarray} \label{eq:canon}
\frac{dr}{dt} &= \frac{\mu \sigma_r^2 \delta}{2r} k (1-\delta/r) \nonumber \\
\frac{dk}{dt} &=   \frac{\mu \sigma_k^2 \gamma}{2r} (r-\delta)^2,
\end{eqnarray} 
where $\sigma_{k,r}^2$ is the variance of the mutational steps. 

\subsubsection*{Trade-off between $k$ and $r$}
We now introduce a strict trade-off between the two traits using a function $r=f(k)$. The evolutionary trajectory now becomes one-dimensional and we view the selection gradient as a function of $k$ only. Throughout this section we will assume that $f(0) = r_0 > \delta$, and that $f(k)$ is decreasing and positive, i.e. it reflects a trade-off (larger $r$ leads to smaller $k$).
The selection gradient now becomes
\begin{equation}\label{eq:grad}
D(k)= \frac{\partial S(k,\hat{k})}{\partial \hat{k}}|_{k = \hat{k}} = \delta \frac{f(k)}{f'(k)} + \frac{\gamma}{k}(f(k)-\delta).
\end{equation}
A rest point of the evolutionary dynamics, known as a singular strategy, is a point $k^{\star}$ such that $D(k^\star)= 0$, i.e. the selection gradient vanishes. 

If we look at \eqref{eq:grad} we notice that since $f$ is decreasing we have $f' < 0$, and the first term is negative. For small $k$ the second term is large and positive, but decays approximately as $f(k)/k$. If there is a $k^{\star} > 0$ such that $D(k^{\star})=0$ then it follows that $\delta / f'(k^{\star}) = -\gamma/ k^{\star} (1-\delta/f(k^{\star}))$. Since the lhs of this equation is less than zero then so must the rhs, and from this we can conclude that $r^\star=f(k^{\star}) > \delta$, i.e. the proliferation rate at the rest point is larger than the death rate (if the converse were true the rest point would be of little interest since the population would go extinct).

A singular strategy $k^{\star}$ is evolutionarily stable and uninvadable if \cite{brannstrom2013}
\begin{equation}\label{eq:ess}
\frac{\partial^2 F(k,\hat{k})}{\partial \hat{k}^2}|_{k = k^{\star}} = \frac{\delta f''}{f} + \frac{2\gamma}{k^{\star}}(1-\delta/f)(f'-\frac{\gamma f}{k^{\star}}) < 0.
\end{equation}
We note that since the second term is always negative the inequality is guaranteed to hold if $f''(k) \leq 0$, i.e. for any concave trade-off function (but could still hold for particular choices of convex $f$). 

For a point to serve as an end point of the evolutionary dynamics it needs to be reachable by the dynamics. This property is called convergence stability and is achieved if
\begin{equation}\label{eq:cs}
\frac{\partial^2 F(k,\hat{k})}{\partial k^2}|_{k = k^{\star}} > \frac{\partial^2 F(k,\hat{k})}{\partial \hat{k}^2}|_{k = k^{\star}}.
\end{equation}

For the particular choice of trade-off we considered $r=f(k)=r_0 e^{-ak}$, the singular strategy is given by 
\begin{equation}
k^{\star} = \frac{W(\frac{e^\gamma \gamma r_0}{\delta}) - \gamma}{a}
\end{equation}
where $W(\cdot)$ is the Lambert $W$ function. We have confirmed numerically that this strategy is both evolutionary and convergence stable.

\clearpage

\setcounter{figure}{0}
\makeatletter 
\renewcommand{\thefigure}{S\@arabic\c@figure}
\makeatother
\begin{figure*}[tbh]
\begin{center}
\includegraphics[width=15cm]{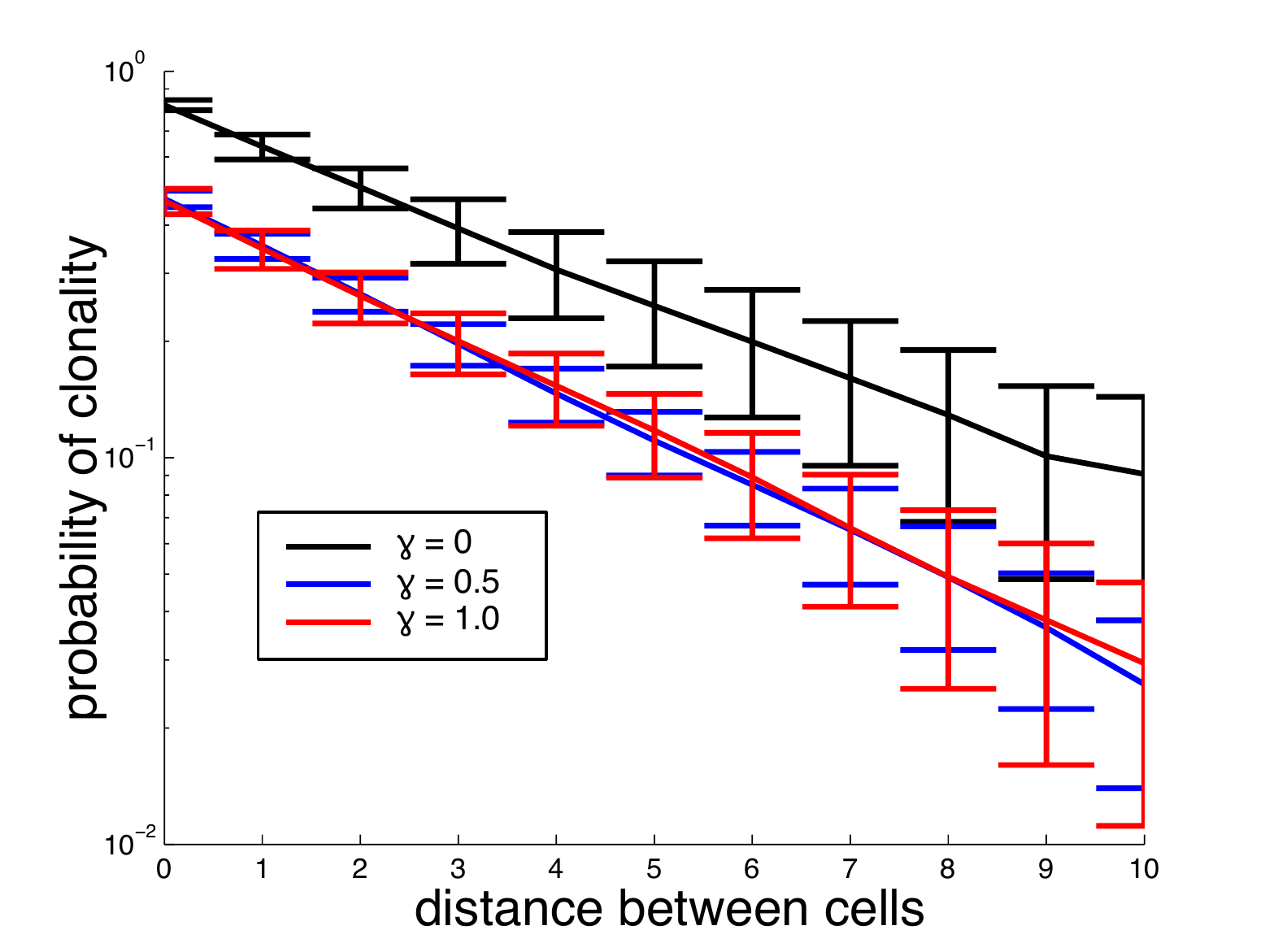}
\caption{\label{fig:S1}The probability that two randomly chosen cells belong to the same subclone as a function of the distance between the cells. The straights lines in the semilog-plot suggests that the clonality scales as $p \sim e^{-\alpha d}$, where $d$ is the distance between the cells. The data was collected from the individual-based model with mutation rate $\mu = 10^{-2}$, death rate $\delta = 0.1$, trade-off function $r=e^{-0.1 k}$ after 4000 time steps. The results were averaged across all cells and 10 independent realisations.} 
\end{center}
\end{figure*}

\begin{figure*}[tbh]
\begin{center}
\includegraphics[width=15cm]{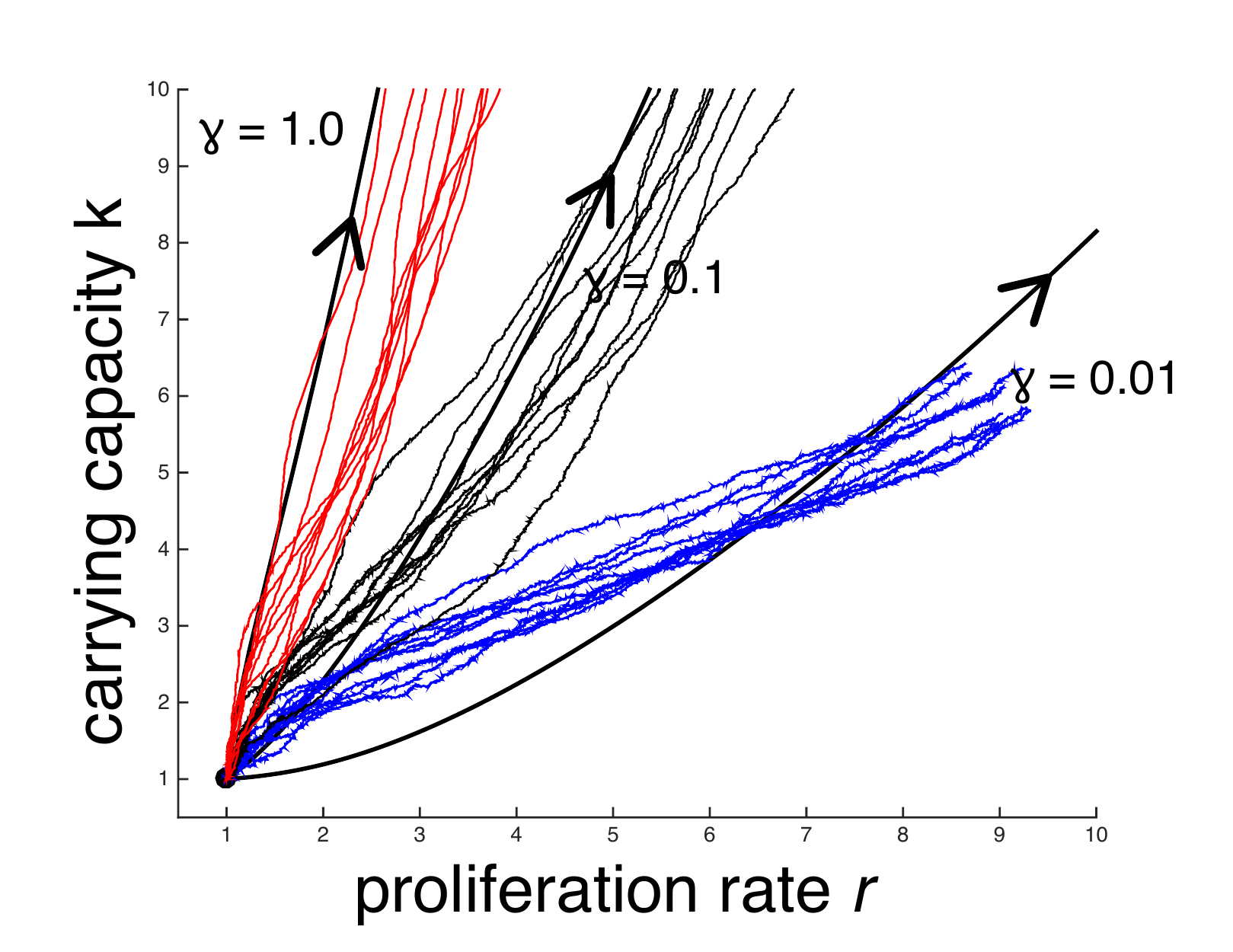}
\caption{\label{fig:S2}Trait substitution sequences when the number of lattice size equals $L=1000$ (coloured curves) and solutions to the canonical equation (10). When the system size is larger the amount of drift is considerably smaller, but is also evident that the canonical equation is less accurate for small values of $\gamma$.} 
\end{center}
\end{figure*}

\begin{figure*}[tbh]
\begin{center}
\includegraphics[width=15cm]{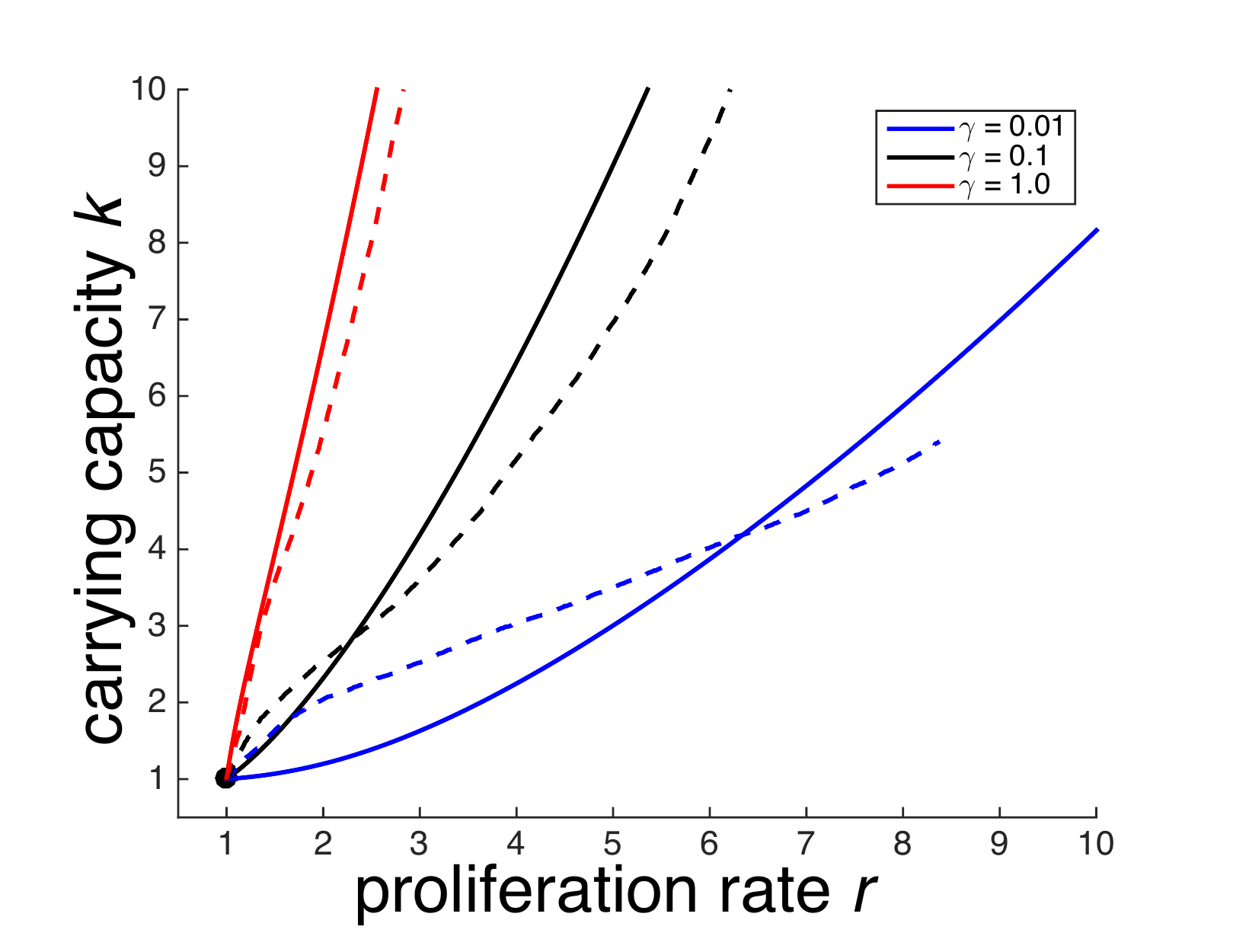}
\caption{\label{fig:S2}Trait substitution curves averaged over 100 simulations (dashed curves) and solutions to the canonical equation (solid curves) for $\gamma=0.01$, 0.1 and 1.0. The deviations are due to small populations size, structured population and a relatively large mutation rate, which leads to an overlap between ecological and evolutionary dynamics. Despite this the canonical equation still gives an approximation of how the direction of selection is influenced by the specificity $\gamma$.} 
\end{center}
\end{figure*}

\end{document}